\documentclass[preprint,aps,pre]{revtex4}
\usepackage[T1]{fontenc}
\usepackage[utf8]{inputenc}
\usepackage{amsmath}
\usepackage{amssymb}
\usepackage{graphicx}
\usepackage[normalem]{ulem}
\usepackage{xcolor}

\begin{document}

\title{Determination of the recombination coefficient in electrolytic solutions from impedance spectroscopy measurements }

\author{G. Barbero$^{1,2}$, N.G. Fytas$^{3}$, I. Lelidis$^{4,*}$}

\affiliation{$^{1}$Dipartimento di Scienza Applicata e Tecnologia del Politecnico di Torino, Corso Duca degli Abruzzi 24, 10129 Torino, Italy\\
$^{2}$National Research Nuclear University MEPhI (Moscow Engineering
Physics Institute), Kashirskoye shosse 31, 115409 Moscow, Russian
Federation\\
$^{3}$ Centre for Fluid and Complex Systems, Coventry University, Coventry CV1 5FB, United Kingdom\\
$^{4}$ Faculty of Physics, National and Kapodistrian University of Athens,\\
Panepistimiopolis, Zografos, Athens 157 84, Greece
\footnote{corresponding author, e-mail: ilelidis@phys.uoa.gr}}
\author{J.V. da Silva Andrade$^5$, F.C.M. Freire$^5$, A.J. Santana$^5$}
\affiliation{$^5$ Departamento de F{\'i}sica, Universidade Estadual de Maring\'a, Avenida Colombo, 5790, 87020-900 Maring\'a, Paran\'a, Brazil}

\date{\today}

\begin{abstract}
	
The dependence of the actual/effective ionic concentration on the nominal salt concentration is experimentally investigated for aqueous solutions of NaCl and KCl. The actual ionic bulk density is determined by means of the impedance spectroscopy technique using two approaches for the data analysis: A first approach scrutinizing the real part's plateau of the cell's electric impedance spectrum versus the bulk density of dopant when the impedance is written in the series representation. A second one via  extraction of the ionic density from simultaneously fitting of the real and imaginary parts of the impedance spectrum by using the Poisson-Nernst-Planck model. The validity of this procedure reposes on the hypothesis that both the diffusion coefficient and the dielectric constant are independent from the salt concentration in the explored range of concentrations. Our experimental results indicate that a first-order reaction describes well the dissociation phenomenon. The dissociation-association ratio of the reaction has been calculated. Deviations from linearity appear at rather low concentrations of the order of $\mathrm{mM}$.
\end{abstract}

\maketitle

\section{Introduction}
\label{sec:1}

The impedance spectroscopy technique is a powerful method to characterize, from the dielectric point of view, solid and liquid materials, by measuring the electric impedance of a sample versus the frequency of an applied electric signal of small enough amplitude. The theoretical analysis of the response of the sample, in the frequency range where the ions present in the medium contribute to the electric response of the system under consideration, is made by the Poisson-Nernst-Planck (PNP) model \cite{mac,PNP}. The PNP model is based on the continuity equations for anions and cations and on the equation of Poisson relating the actual potential to the bulk ionic charge density. Several versions of this model have been proposed along the years considering different boundary conditions and physical systems \cite{cj,bio1,bio2,cement,mac1,serghei1,serghei2}.

The spectra of the impedance's real and imaginary parts in the series representation in the low frequency regime (smaller than one hundred Hz) strongly depend on the physical properties of the electrodes. On contrary, for frequencies larger than approximately one kHz, the electric response of the cell is independent of the electrodes, which can be considered as blocking. In this frequency range, the real part of the impedance presents a large plateau whose amplitude depends only on the diffusion coefficient, $D$, and the bulk density of ions, $n$, in thermodynamic equilibrium~\cite{alai}. From the analysis of this part of the spectrum it becomes possible to obtain information about $D$ and $n$. Taking into account that the diffusion coefficient, in the limit of small salt concentrations, is independent of the solute's concentration~\cite{atkins}, $D$ can be considered as a known quantity, and $n$ can be derived from this part of the spectrum. Alternatively, $D$ and $n$ can be calculated by fitting the real and imaginary parts of the impedance in order to validate the plateau method.

Recently, we have investigated the concentration dependence of the capacitive to inductive transition observed in electrolytic solutions~\cite{JML}, related to the presence of a parasite inductance connected to the connection cables. In that particular case, the parasite inductance was found to be responsible for a cusp in the modulus of reactance, appearing at a well defined frequency, depending on the salt concentration present in the solution.

Here, we use the same system to investigate the dissociation of NaCl and KCl in water and, in particular, the dependence of the recombination coefficient on the salt concentration. To this end, we calculate the actual/effective density of ions in the cell and compare it to the added salt concentration in the water. For the data analysis, we use the plateau method to obtain the actual ion's density. In addition, the effective ionic concentration is calculated by fitting the real and imaginary parts of the impedance.

The remainder of this paper is organized as follows: In Sec.~\ref{sec:2} we outline the Poisson-Nernst-Planck model used for the analysis of experimental data and provide some useful formulas for deriving the ionic bulk density at the impedance level. In Sec.~\ref{sec:3} we present our experimental results focusing on the dependence of recombination coefficient on salt concentration. This paper ends with summary and conclusions in Sec.~\ref{sec:4}.

\section{Theoretical Framework}
\label{sec:2}

\subsection{Impedance}
\label{subsec:2a}

In the Poisson-Nernst-Planck model's framework, we have shown that the electric impedance $Z$ of an electrolytic cell in the shape of a slab of thickness $d$ and surface area $S$ limited by blocking electrodes, is given by~\cite{alai}
\begin{equation}
\label{eq:1}Z=\frac{2}{i \omega \varepsilon \beta^2 S}\left\{\frac{1}{\lambda^2 \beta}\,\tanh\left(\beta \frac{d}{2}\right)+i\,\frac{\omega d}{2 D}\right\}.
\end{equation}
In Eq.~(\ref{eq:1}) above, $\omega$ denotes the circular frequency of the applied external field, $\varepsilon$ the dielectric constant of the liquid free of ions, $D$ the diffusion coefficient of the ions in liquid,
$\lambda=\sqrt{\varepsilon k_BT/(2 n q^2)}$ the Debye length, and
\begin{equation}
\label{eq:2}\beta=\frac{1}{\lambda}\,\sqrt{1+i \omega \,\frac{\lambda^2}{D}}.
\end{equation}
In the expression defining $\lambda$, $k_{\rm B}$ is the Boltzmann constant, $T$ the absolute temperature, $q$ the ionic charge, and $n$ the bulk density of ions in thermodynamic equilibrium. Equation~(\ref{eq:1}) holds in the case where cations and anions have the same electrical charge and diffusion coefficient, as happens for NaCl and KCl. A generalization of the analysis in cases where ions have different valence and diffusion coefficients and the electrodes are not blocking has been presented in Ref.~\cite{antonova}. Yet, for the case under consideration the simple analysis relevant to the monovalent ions, with the same diffusion coefficient, is enough for the analysis of the experimental data.

Introducing the Debye relaxation circular frequency $\omega_{\rm D}=D/\lambda^2$, Eq.~(\ref{eq:1}) can be rewritten as
\begin{equation}
\label{eq:3}Z=\frac{i \omega d+2 \omega_{\rm D} \tanh (\beta d/2)}{i \omega \varepsilon \beta (\omega_{\rm D}+ i \omega)S}.
\end{equation}
If in the circuit a parasite inductance $L$ is also present, then the  reactive contribution $i \omega L$ appears as well, and Eq.~(\ref{eq:3}) needs to be reformulated as  folows
\begin{equation}
\label{eq:4}Z=\frac{i \omega d+2 \omega_{\rm D} \tanh (\beta d/2)}{i \omega \varepsilon \beta (\omega_{\rm D}+ i \omega)S}+i \omega L.
\end{equation}

In this form the expression of the electric impedance has been used recently to investigate the dependence of the capacitive-inductive transition on the ionic density~\cite{JML}. Defining
\begin{equation}
\label{eq:5}M=\frac{d}{2 \lambda},\quad R_{\rm p}=\frac{d}{\varepsilon \omega_{\rm D} S},\quad \Omega=\frac{\omega}{\omega_{\rm D}},\quad{\rm and}\quad {\cal L}=\omega_{\rm D} L,
\end{equation}
equation~(\ref{eq:4}) can be put in the form
\begin{equation}
\label{eq:6} Z=R_{\rm p}\,\,\frac{i \Omega M \sqrt{1+i \Omega}+\tanh(M \sqrt{1+ i \Omega})}{i \Omega M (1+i \Omega)^{3/2}}+i \Omega {\cal L}.
\end{equation}
Usually $\lambda\ll d$, and hence $M\gg 1$. In this framework Eq.~(\ref{eq:6}) is very well approximated by
\begin{equation}
\label{eq:7} Z=R_{\rm p}\,\,\frac{1+i \Omega M \sqrt{1+i \Omega}}{i \Omega M (1+i \Omega)^{3/2}}+i \Omega {\cal L}.
\end{equation}

The quantity indicated by $R_{\rm p}$ coincides to the value of the plateau of the spectrum of $R = R(\omega)$~\cite{alai}. As it follows from its definition, $R_{\rm p}$ depends on both $\lambda$ and $D$. However, for aqueous solutions with concentrations as those investigated in the current work, the diffusion coefficient depends only weakly on the concentration and it is well approximated by that of the diluted solution~\cite{atkins}. From the measured value of the resistance of the plateau $R_p$, for a given nominal concentration $n_0$, it is then possible to evaluate the actual ionic density $n$.

For circular frequency very small with respect to $\omega_{\rm D}$, yet still high enough to neglect the contribution of electrodes to the impedance, such that $\Omega\ll 1$, we get from Eq.~(\ref{eq:7})
\begin{equation}
\label{eq:8} Z\sim R_{\rm p}-i\left(\frac{R_{\rm p}}{M \Omega}+(R_{\rm p}-{\cal L})\Omega\right).
\end{equation}
From this expression it follows that in this frequency regime the modulus of the reactance has a minimum for the reduced circular frequency
\begin{equation}
\label{eq:9} \Omega_{\rm m}=\frac{1}{\sqrt{M(1-{\cal L}/R_{\rm p})}},
\end{equation}
whose value is
\begin{equation}
\label{eq:10} |X(\Omega_{\rm m})|=2\,\frac{R_{\rm p}}{\sqrt{M}}\,\sqrt{1-\frac{{\cal L}}{R_{\rm p}}}.
\end{equation}
These relations indicate that, changing $R_{\rm p}$ is also changing the minimum of the reactance modulus and its position, as it is experimentally observed.

\subsection{Recombination-dissociation phenomenon}
\label{subsec:2b}

In order to derive information on the recombination-dissociation phenomenon, we assume that the decomposition of a salt in water can be described by a chemical reaction whose order depends on the ions generated. In the case where the produced ions have the same valence, the reaction is of first order, and the neutral molecule $\rm A$ dissociates in two monovalent ions $\rm B^+$ and $\rm C^-$, that is, $\rm A = \rm B^{+}+\rm C^{-}$, according to Refs.~\cite{derfel,ioannis,wang,JML}. For an infinite sample ($d\to \infty$) in the absence of any external power supply, the system is at thermodynamic equilibrium, and the bulk density of neutral, positive, and negative species are position-independent. If we denote by $n_0$, $n_{\rm n}$, and $n$ the bulk densities of salt molecules that can dissociate, neutral, and charged particles respectively, and by $k_{\rm d}$ and $k_{\rm a}$ the constants of dissociation and recombination, we arrive at
\begin{equation}
\label{eq:11} n+n_{\rm n}=n_0,\quad{\rm and}\quad  k_{\rm d}\,n_{\rm n}=k_a\,n^2.
\end{equation}
It follows that the bulk density of ions is given in thermodynamic equilibrium by
\begin{equation}
\label{eq:12} n=-\,\frac{k_{\rm d}}{2 k_{\rm a}}+\sqrt{\left(\frac{k_{\rm d}}{2 k_{\rm a}}\right)^2+\frac{k_{\rm d}}{k_{\rm a}}\,n_0}.
\end{equation}
The degree of dissociation defined as $\rho=n/n_0$ follows then as
\begin{equation}
\label{eq:13} \rho=-\frac{\kappa}{2}+\sqrt{\left(\frac{\kappa}{2}\right)^2+\kappa},
\end{equation}
where $\kappa=k_{\rm d}/(k_{\rm a} n_0)$. From the measured values of $n$, and via the nominal concentration  $n_0$, it is possible to determine the ratio $k_{\rm d}/k_{\rm  a}$ of the reaction.

\section{Experimental Part and Analysis}
\label{sec:3}

We measured the impedance spectra of saline solutions composed of (i) potassium cloride (KCl) and (ii) sodium chloride (NaCl), both in Milli-Q water. The concentration of solutions in NaCl or KCl salt varied from $3\times 10^{22}/\mathrm{m^3}$ up to $1.42\times 10^{25}/\mathrm{m^3}$, see Tab.~\ref{tab:1}. The full description of the saline solution preparation and impedance spectroscopy measurement is given in Ref.~\cite{JML}. Impedance measurements were performed by applying a harmonic potential of 25 mV in amplitude between the electrodes of the cell containing the electrolyte. Linearity has been tested by varying the amplitude of the applied potential. In what follows we present the impedance spectra of three KCl solutions in order to describe the results and manifest the analysis method. All impedance spectra are given in the Appendix.

\begin{table}
    \centering
   \caption{\label{tab:1} Solution number (column 1) and the corresponding concentration value for each salt (column 2 in $\mathrm{m^{-3}}$, column 3 in  $\mathrm{mol/L}$). Solution-0 is pure Milli-Q water.}
	
\begin{tabular}{ccc}
\hline
Solution & Concentration $[\times 10^{24}\,\mathrm{/m^3}]$ & Concentration{[}$\times 10^{-3}\mathrm{mol/L}${]}%\cline{3-6}
                          \\ \hline\hline
0                     &	Milli-Q Water &	Milli-Q Water \\ \hline
1                     &	0.03      &	0.05    \\ \hline
2                     &	0.07      &  0.11    \\ \hline
3                     &	0.10      &  0.16   	\\ \hline
4                     &	0.13      &  0.22    \\ \hline
5                     &	0.16      &  0.27    \\ \hline
6                     &	0.31      &  0.51    \\ \hline
7                     &	0.45       & 0.75     \\ \hline
8                     &	0.59      &  0.98     \\ \hline
9                     &	0.73      & 1.21    \\ \hline
10                    &	0.87      & 1.44   \\ \hline
11                    &	2.02       & 3.35     \\ \hline
12                    &	3.12      & 5.18     \\ \hline
13                    &	4.19       & 6.95    \\ \hline
14                    &	5.21       & 8.65     \\ \hline
15                    &	6.20      & 10.29   \\ \hline
16                    &	14.42     & 23.94   \\   \hline
\end{tabular}
\end{table}

\begin{figure}[htb]
	\includegraphics[width=0.45\textwidth]{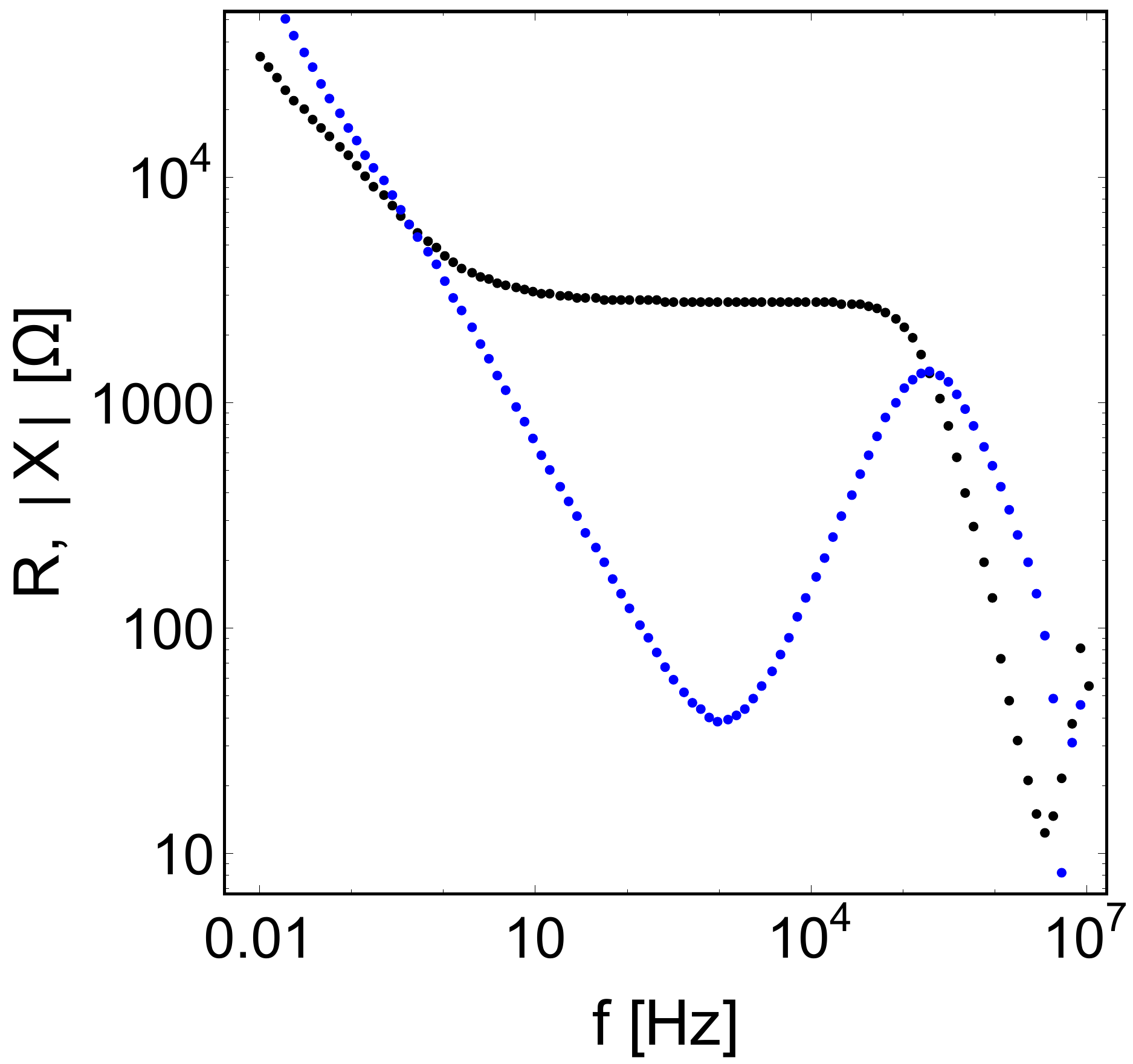}
	\caption{Resistance $R$ (black points) and absolute value of reactance $\left | X \right|$ vs. frequency $f$ for the KCl solution with ionic concentration $0.03\times 10^{24}/\mathrm{m^3}$. The cusp in $\left | X \right|$ is due to the capacitive-inductive transition.}
	\label{RX1}
\end{figure}

\begin{figure}[htb]
	\includegraphics[width=0.45\textwidth]{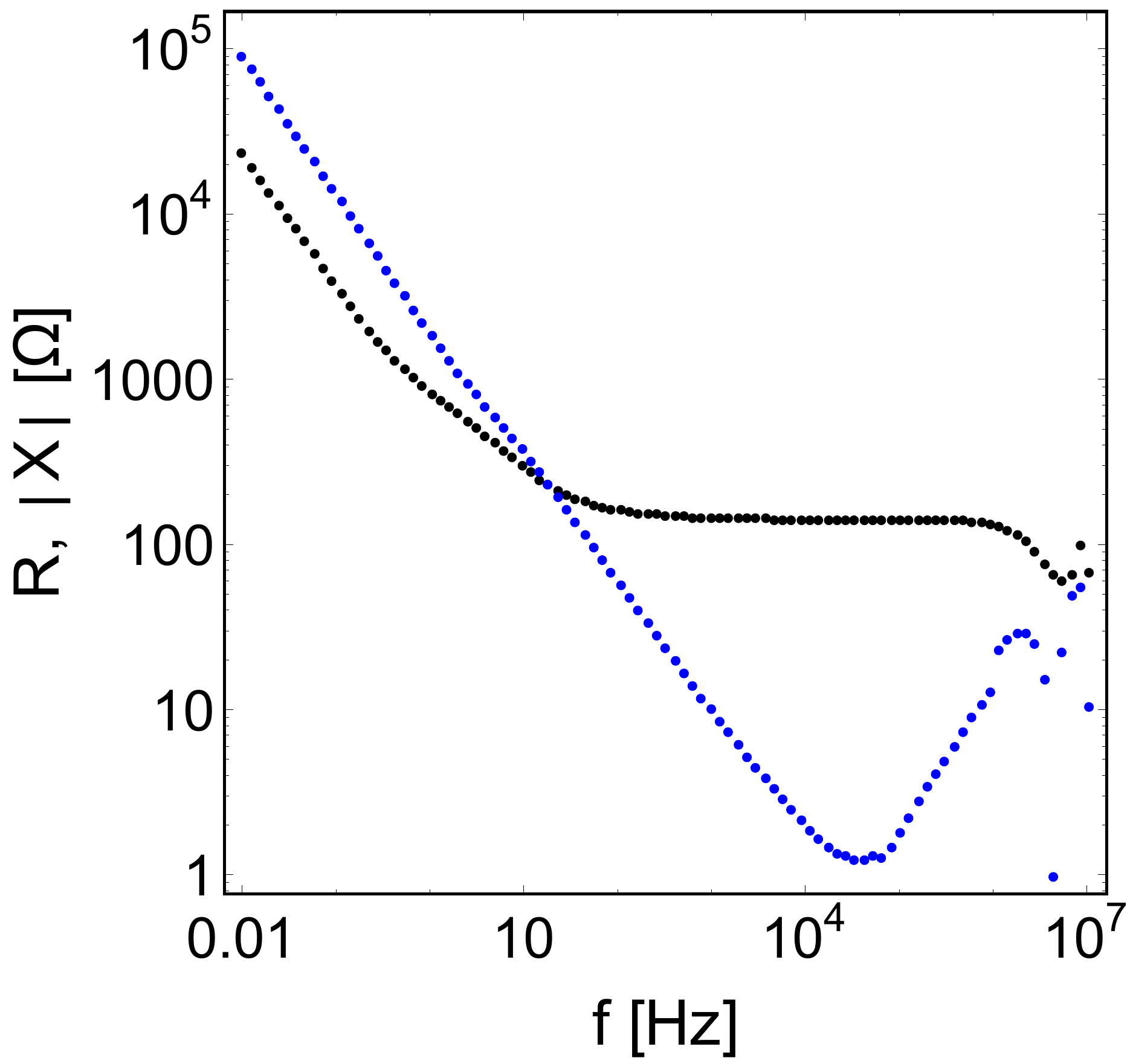}
	\caption{Similar to Fig.~\ref{RX1} but with ionic concentration of $0.73\times 10^{24}/\mathrm{m^3}$. }
	\label{RX9}
\end{figure}

\begin{figure}[htb]
	\includegraphics[width=0.45\textwidth]{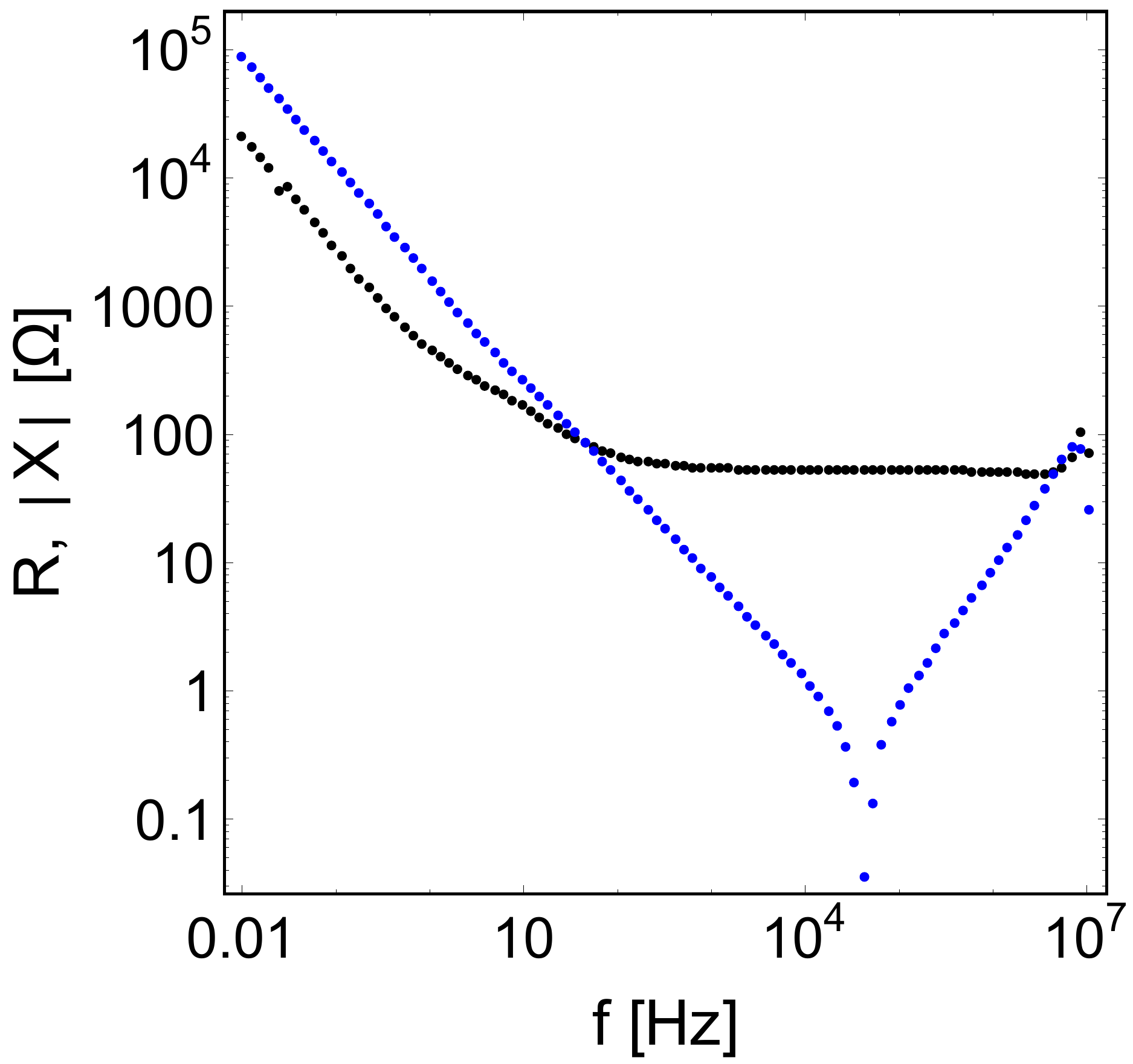}
	\caption{Similar to Fig.~\ref{RX1} but with ionic concentration of  $2.02\times 10^{24}/\mathrm{m^3}$.}
	\label{RX11}
\end{figure}

Figures~\ref{RX1}--\ref{RX11} illustrate the real ($R$) and imaginary ($\left|X\right|$) parts of the impedance for three typical concentrations of KCl (The spectra for NaCl are similar and are not shown here for brevity). The appearance of a cusp in the absolute value of the reactance results stems from a parasitic inductance due to the coaxial cables of the experimental set up~\cite{JML}. This latter inductance was measured equal to $L=2\,\mathrm{\mu H}$~\cite{JML}. The cusp signals a transition from the capacitive to inductive regime~\cite{negative}. The dependence of the cusp's frequency position on the ionic concentration reveals the existence of a critical concentration where this frequency shows a jump of approximately two orders of magnitude, \emph{i.e.}, from the MHz regime to a few tens of KHz~\cite{JML}. 

\begin{figure}[htb]
	\includegraphics[width=0.45\textwidth]{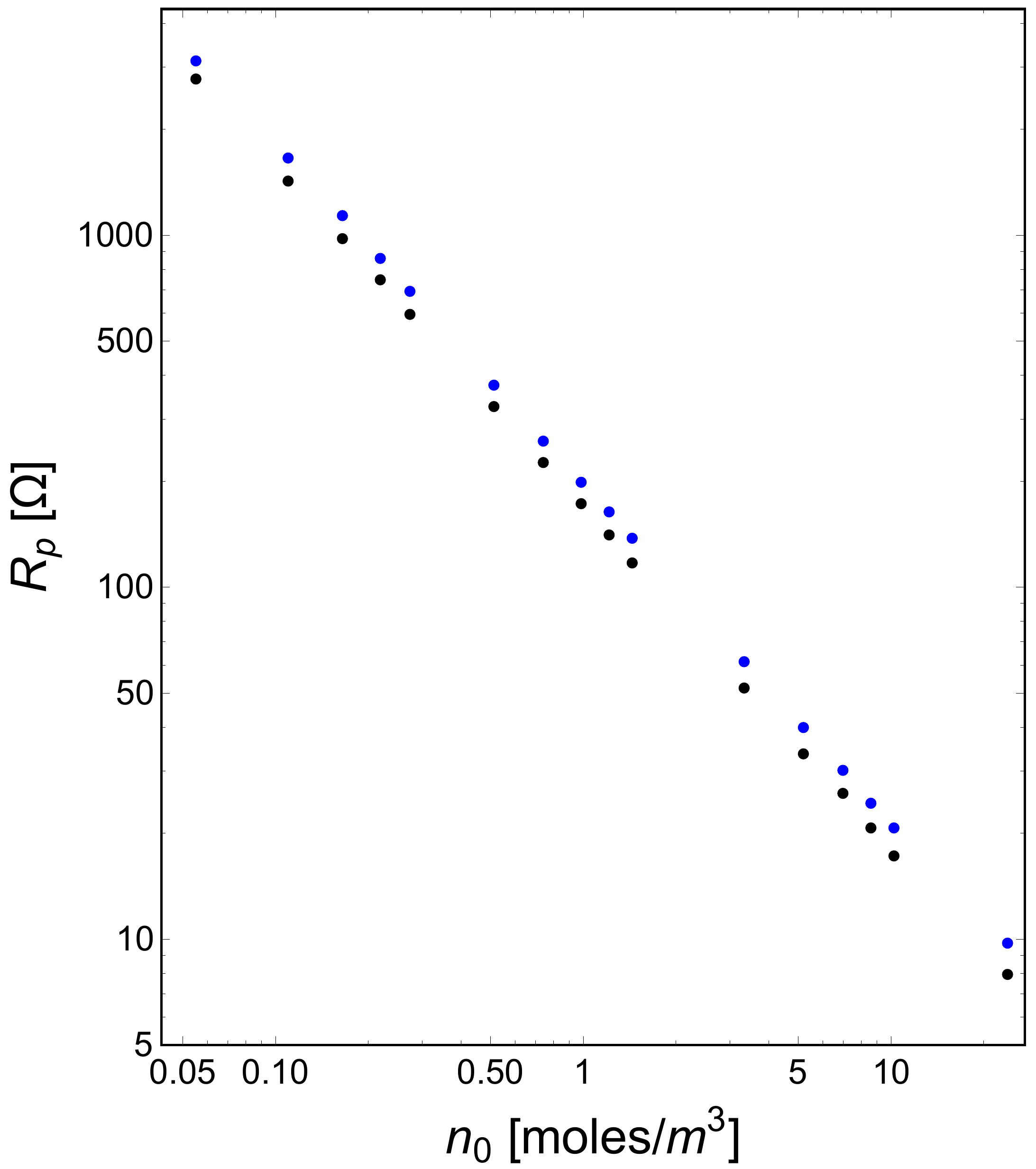}
	\caption{Resistance at the plateau, $R_{\rm p}$, vs. the nominal ion concentration $n_0$. Data for KCl (black points) and NaCl (blue points) are shown. }
	\label{Rp}
\end{figure}

In order to obtain the effective ionic concentration $n$ as function of the nominal one $n_0$ for the various salt solutions and use the obtained results to reckon the dissociation-regeneration coefficient, we follow the protocol of Sec.~\ref{sec:2}. This is based on the experimental value of the resistance at the plateau, $R_{\rm p}$.  These $R_{\rm p}$ values are shown in Fig.~\ref{Rp} for both KCl and NaCl. Note the minor differences in their values which is due to the different diffusion constants.

\begin{figure}[htb]
	\includegraphics[width=0.45\textwidth]{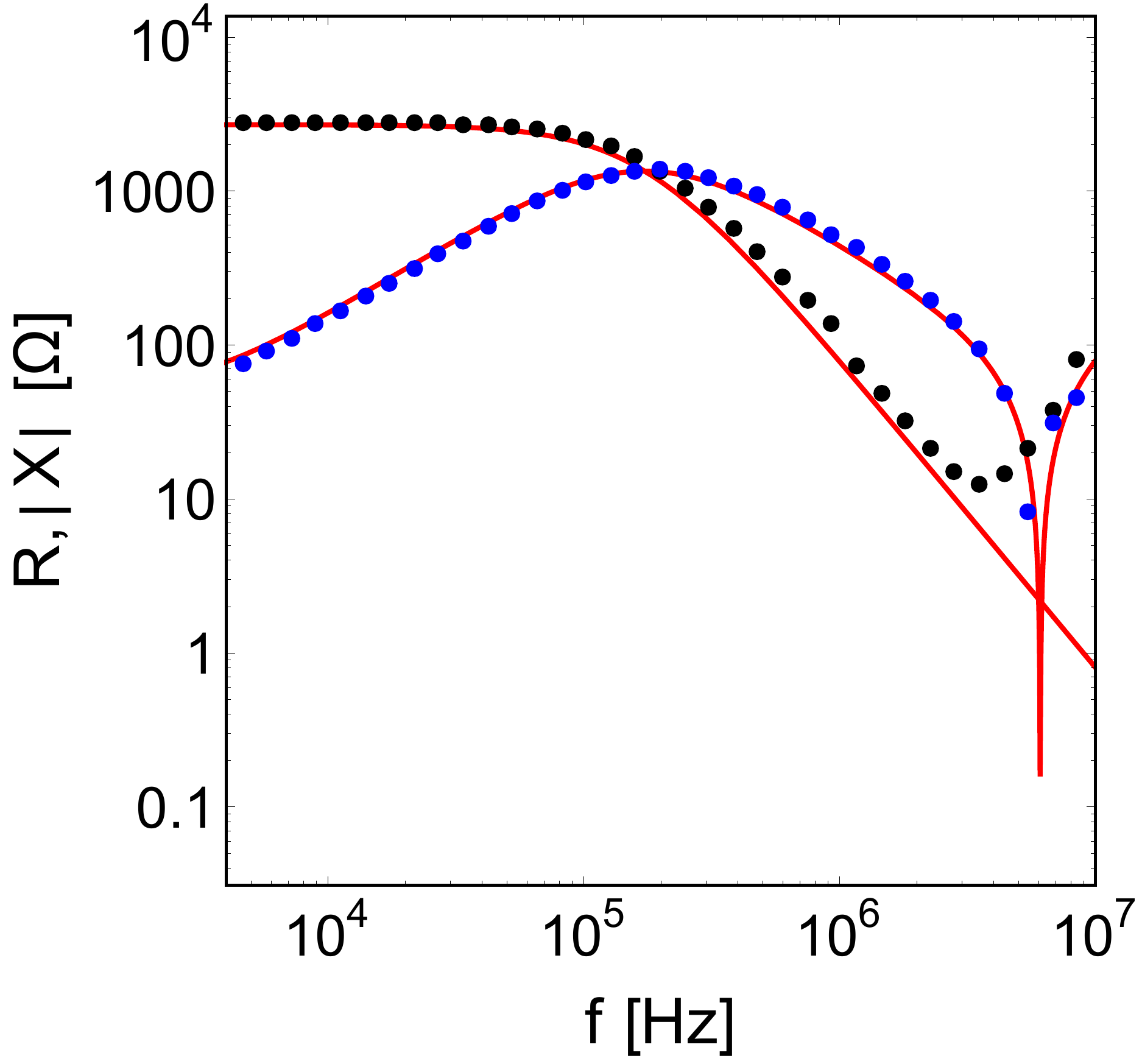}
	\caption{Resistance $R$ (black points) and absolute value of reactance $\left | X \right|$ vs. frequency $f$ for a KCl solution with ionic concentration of $0.03\times 10^{24}/\mathrm{m^3}$, far below the critical concentration for the capacitive-inductive transition. Continuous lines correspond to the best fit of the shown experimental data points using the PNP model.}
	\label{F1}
\end{figure}

\begin{figure}[htb]
	\includegraphics[width=0.45\textwidth]{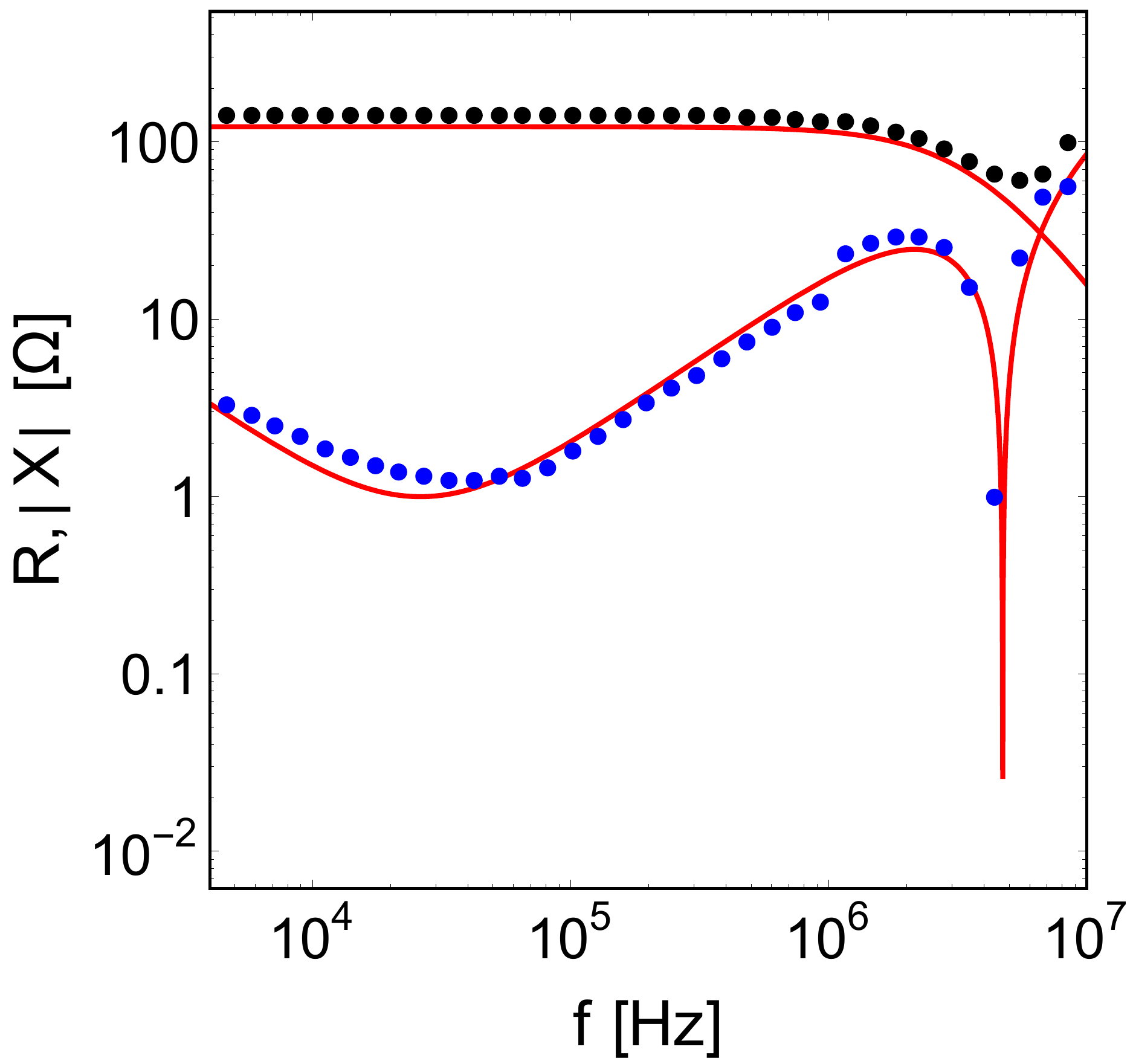}
	\caption{Similar to Fig.~\ref{F1} but here the ionic concentration is taken $0.726\times 10^{24}/\mathrm{m^3}$, close to the critical concentration for the capacitive-inductive transition.}
	\label{F9}
\end{figure}

\begin{figure}[htb]
	\includegraphics[width=0.45\textwidth]{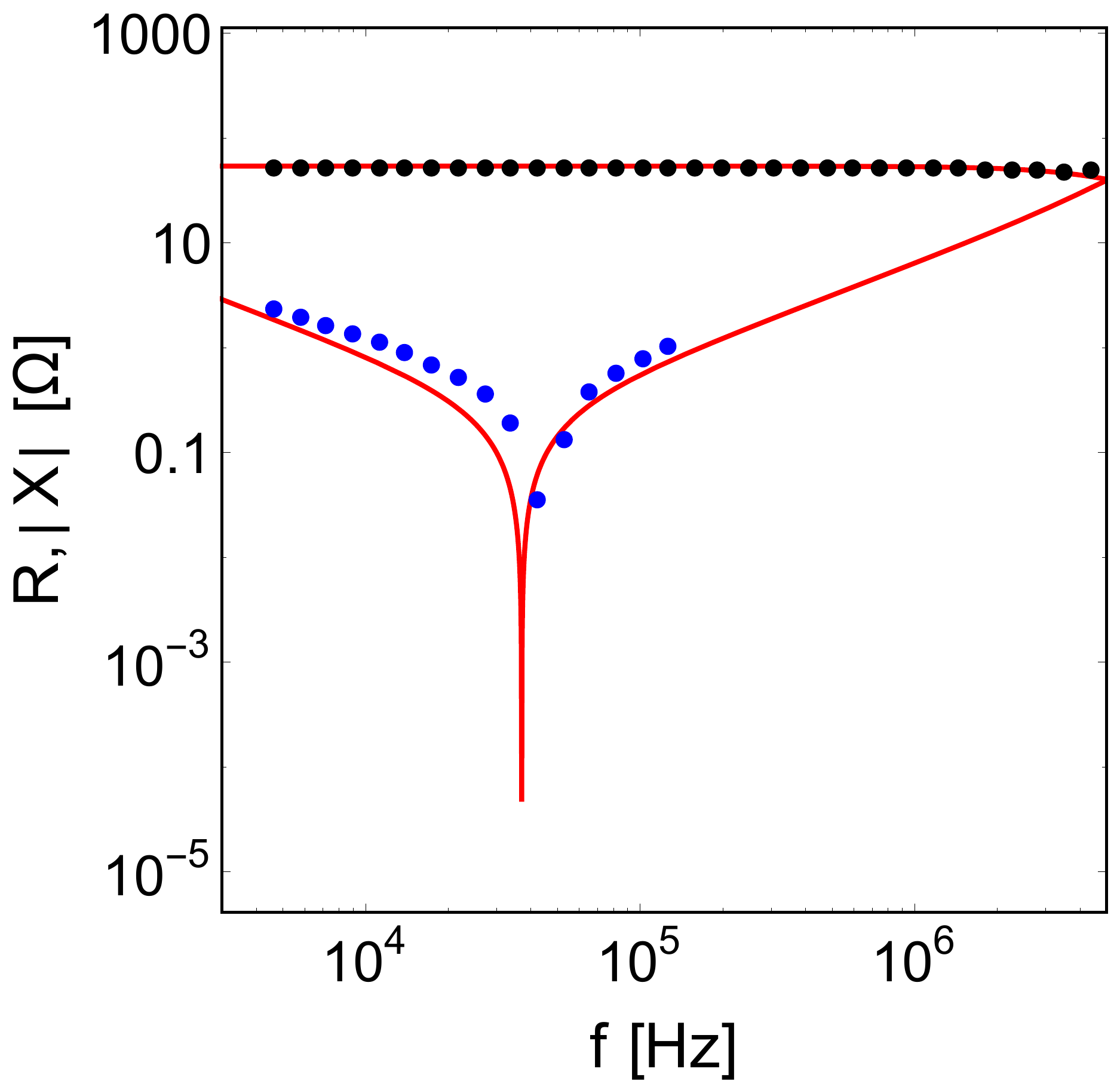}
	\caption{Similar to Fig.~\ref{F1} but here the ionic concentration is taken $2.01\times 10^{24}/\mathrm{m^3}$, much larger than the critical concentration for the capacitive-inductive transition.}
	\label{F11}
\end{figure}

For the numerical calculations the diffusion coefficient at room temperature of KCl is taken equal to $D_{\rm KCl}=2\times 10^{-9}\,\mathrm{m^2/s}$. For NaCl we use $D_{\rm NaCl}=1.6\times 10^{-9}\,\mathrm{m^2/s}$ respectively~\cite{D,D2,diffusion}. Note that in the case of NaCl the diffusion coefficient for anions and cations are different. Nevertheless, in the plateau range free diffusion is important since the ambipolar effect is crucial only at lower frequencies~\cite{ambi}. The ions are monovalent with $q=1.6\times 10^{-19}\,\mathrm{C}$ and the water dielectric constant is taken as $\varepsilon\sim 80\times \varepsilon_0$~\cite{atkins}. The geometrical parameters of the experimental cylindrical cell are: (i) the thickness $d=0.65\times10^{-3}$m, and (ii) the surface area  $S=\pi\times 10^{-4}$m$^2$ of the electrodes. From the expression $R_{\rm p}=k_{\rm B}Td/(2q^2\,S\,D\,n)$, and using the experimental values of the plateau resistance, one may calculate the effective concentration of ions $n$. In order to test the obtained results, we used them to reproduce the resistance and reactance of the cell for each concentration. This procedure indicates that for low concentrations one can reproduce the experimental spectra fairly well in the regime above the zone where boundary conditions influence the sample's response. Yet, the method seems to fail above the critical concentration. Therefore, we tried also to fit simultaneously $R(f)$ and $\left|X(f)\right|$, as shown in Figs.~\ref{F1}--\ref{F11}. Continuous lines are the result of fitting using the PNP model. We fit $R(f)$ and $\left|X(f)\right|$ only in the frequency range above a few kHz in order to avoid complication due to the electrodes. At low concentrations we find results compatible to those obtained via the $R_{\rm p}$ method, while at higher concentrations the obtained results differ by less than $\sim 20\%$.

\begin{figure}[htb]
	\includegraphics[width=0.45\textwidth]{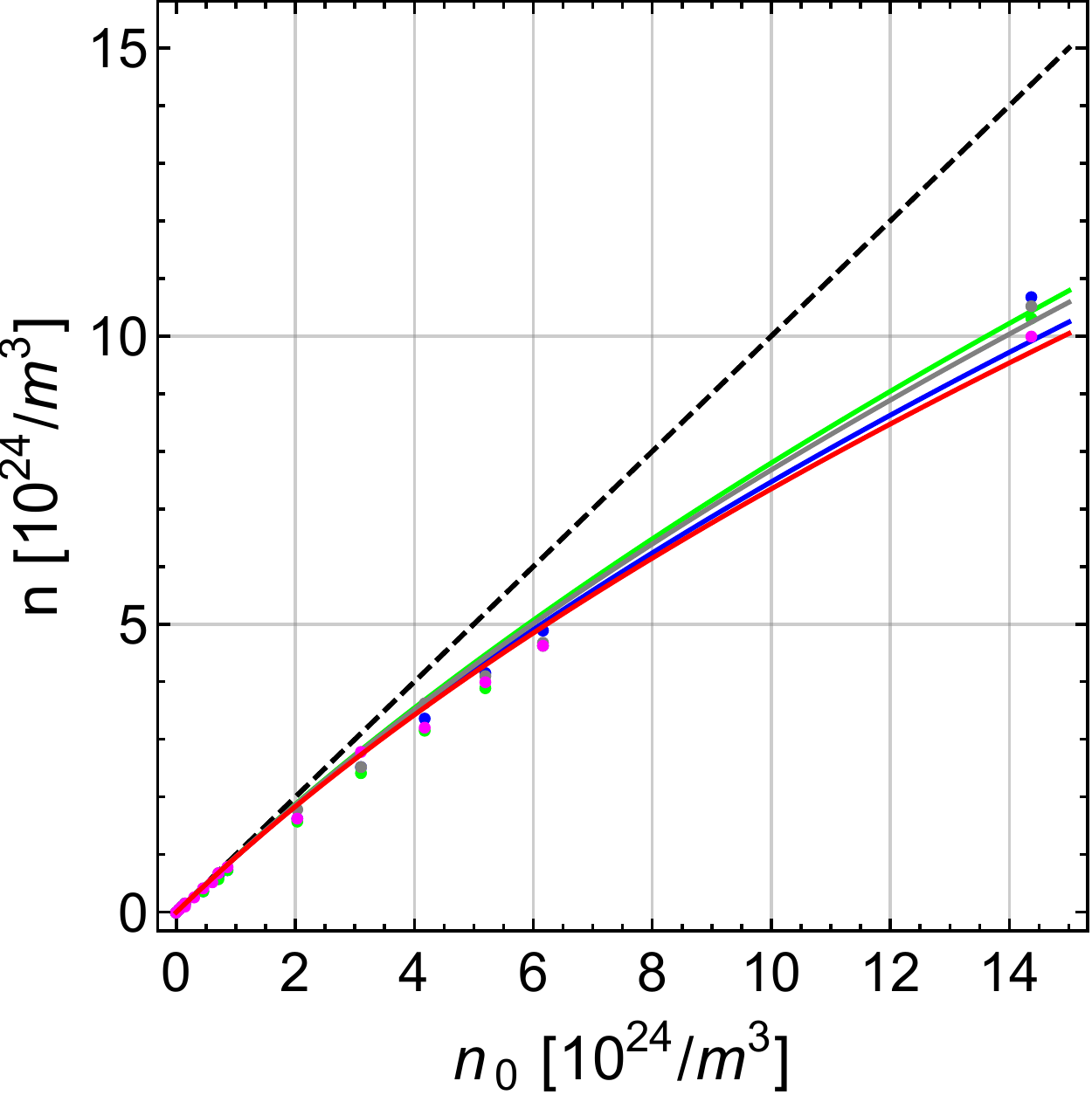}
	\caption{Effective vs. nominal concentration for KCl and NaCl solution in Milli-Q water. The dashed line corresponds to complete dissociation. Points are calculated with (i) the plateau method and (ii) by fitting simultaneously the real and imaginary part of the experimentally measured complex impedance. Continuous lines are best fits for the recombination constant.}
	\label{recomb}
\end{figure}

Finally Fig.~\ref{recomb} shows the effective ionic concentration versus the nominal one. Points are calculated via: (i) the $R_{\rm p}$ method (blue points for NaCl and green ones for KCl) and (ii) fitting $R(f)$ and $\left|X(f)\right|$ (red points for NaCl and gray ones for KCl). Continuous lines show the best fits using Eq.~(\ref{eq:12}). The colors of the continuous line and corresponding points are identical and the dashed line marks complete dissociation. The obtained values for the ratio $k=k_{\rm d}/k_{\rm a}$ are listed here: $k_{\rm NaCl}=2.76\times 10^{25}/\mathrm{m^3}$ and $k_{\rm KCl}=2.54\times 10^{25}/\mathrm{m^3}$ fro the $R_{\rm p}$ method and  $k_{\rm NaCl}=2.21\times 10^{25}/\mathrm{m^3}$ and $k_{\rm KCl}=2.04\times 10^{25}/\mathrm{m^3}$ from the simultaneous fit of $R(f)$ and $\left|X(f)\right|$. Several comments are in the order at this point: \\
(i) Discrepancies among KCl and NaCl effective concentrations are small and could originate from the approximated diffusion constants used in the above analysis. \\
(ii) With increasing concentration of the solution, the association-dissociation effect has to be taken into account even for concentrations of $\sim 10^{-3}\,\mathrm{M}$. \\
(iii) Other contributions resulting from at distance interactions among ions and among ions and solvent molecules might be important especially at higher concentrations than the ones considered here~\cite{H2O,marcus}.

Closing, we would like to mention that we have limited our analysis at frequencies above 10 kHz where the influence of electrodes' properties on the electric response of the cell to the external harmonic excitation is negligible. Moreover, in this frequency regime the ambipolar diffusion is not significant. Adding a Warburg element in order to 
take into account the role of electrodes, in series to the impedance evaluated by means of the 
PNP-model, it is possible to fit the experimental data over the entire 
frequency range. Warburg impedance~\cite{macdo,tribollet} has the functional form 
$Z_{\rm W}=R_{\rm W}\/i(\omega/\omega_W)^\gamma$ where $R_{\rm W}$ is a real observable having the dimension of a resistance, $\omega_{\rm W}$ a typical circular frequency, and $\gamma<1$, are parameters of the model, depending on 
the state of electrodes and the interface electrode-solution. 
However, several phenomena contribute to $Z_{\rm W}$, as the adsorption,
ambipolar diffusion, roughness of the surface, and others. Considering $R_{\rm W}$, $\omega_{\rm W}$, and $\gamma$ as fitting parameters it is possible to obtain a good quality fit, still without being possible to decipher the physical origin of these parameters. For this reason we limited our analysis at high enough frequencies, as described above.

% a good fitting does not always implies a better understanding of the system's physics

\section{Conclusions}
\label{sec:4}

To summarize, we have calculated the effective concentration of ions in NaCl and KCl solutions in water as a function of the nominal salt concentration by the plateau method and by fitting the measured impedance using the PNP model. As shown both methods gave compatible results giving credit to our two-fold analysis. Assuming that the salt dissociation is described by a first-order reaction we calculated the reaction quotient. The analysis has been done in the approximation of constant diffusion coefficient and dielectric constant. It would be interesting to compare results of the above described method for evaluation of the effective ion concentration with measurements on the same samples by other experimental methods, such as conductivity \cite{bockris} and spectroscopic ones \cite{spectro1,spectro2,rev1,ftir}.

{\bf Acknowledgements}

G.B. acknowledges partial support by the MEPhI Academic
Excellence Project.
A.J.S and J.V.S.A. acknowledge financial support from the Brazilian Agency e Conselho Nacional de Desenvolvimento Cient\'{\i}fico e Tecnol\'{\o}gico (CNPq) and the Coordena\c{c}\~{a}o de Aperfei\c{c}oamento de Pessoal de N\'{\i}vel Superior - Brasil (CAPES) - Finance Code 001.

\newpage

\clearpage

\appendix{\textbf{Appendix}}

\begin{figure}[htb]
	\includegraphics[width=0.75\textwidth]{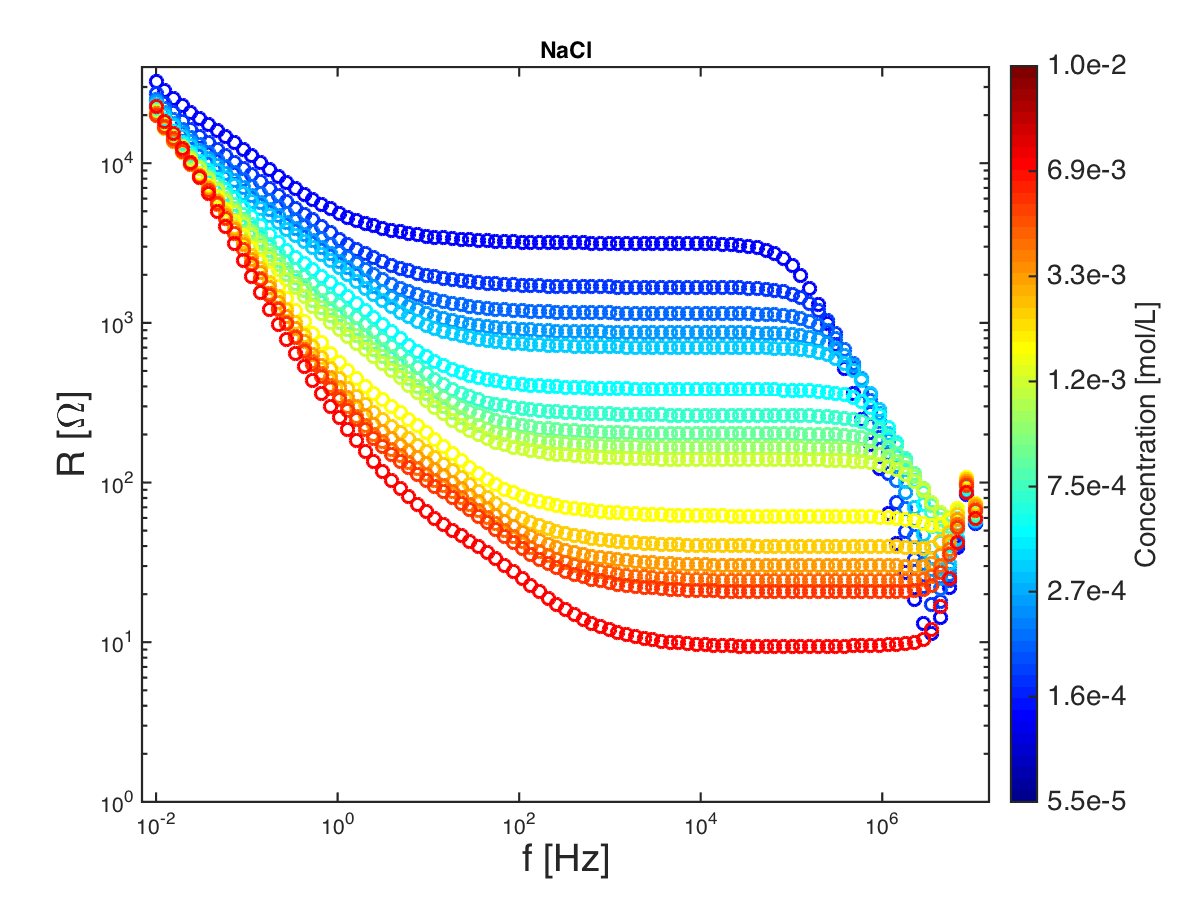}
		\caption{Resistance ($R$) vs. frequency ($f$) for the NaCl solutions in a double logarithmic scale. The solution number is increasing moving from the blue to the red curve. The lowest curve (red circles) corresponds to the highest concentration, that is, to the solution 16 (see Tab.~\ref{tab:1}). The curve in blue circles at the apex of the stack corresponds to the Milli-Q water (solution 0 in Tab.~\ref{tab:1}).}
		\label{NaClR}
\end{figure}

\begin{figure}[htb]
	\includegraphics[width=0.75\textwidth]{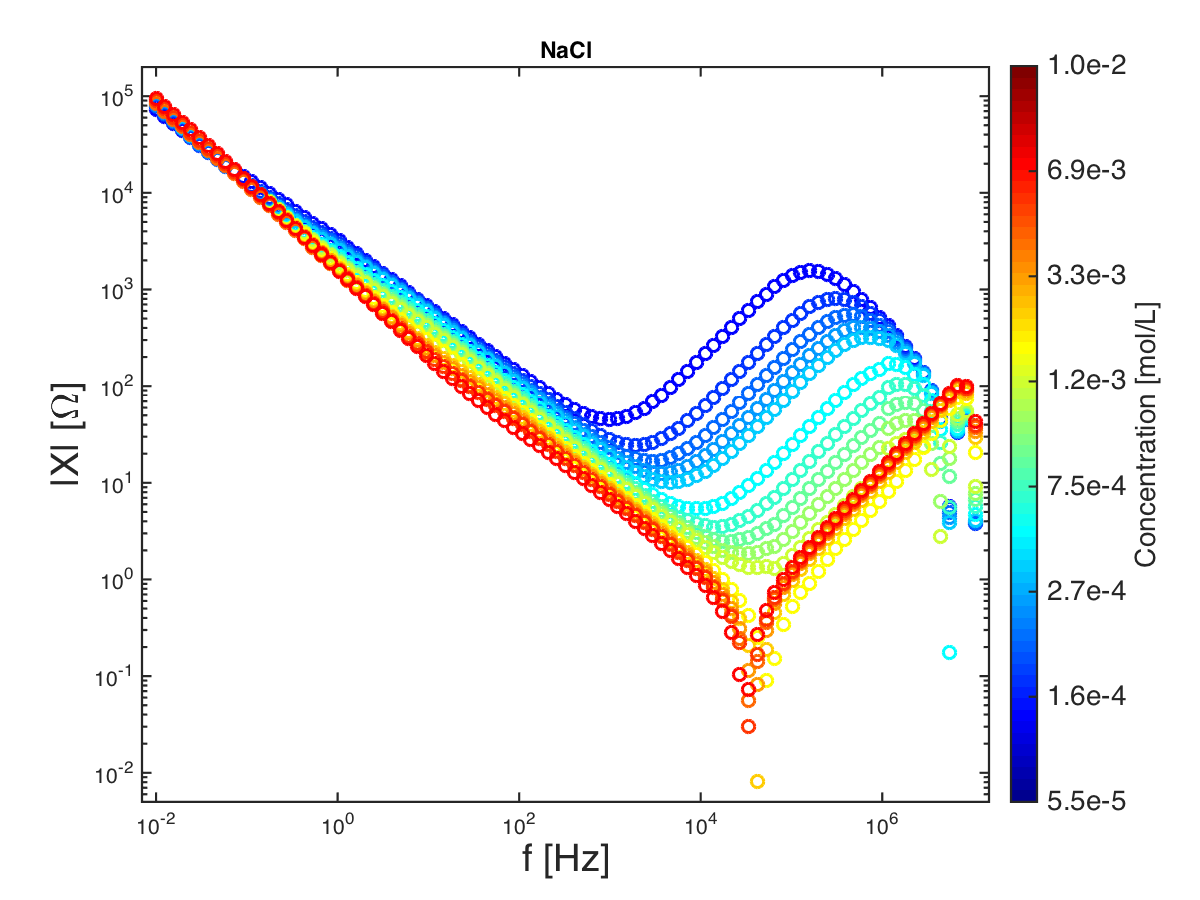}
		\caption{Absolute value of the reactance ($\left|X\right|$) vs. frequency ($f$) for the NaCl solutions in a double logarithmic scale. Symbols and concentrations as in Fig.~\ref{NaClR}.}
		\label{NaClX}
\end{figure}

\begin{figure}[htb]
	\includegraphics[width=0.75\textwidth]{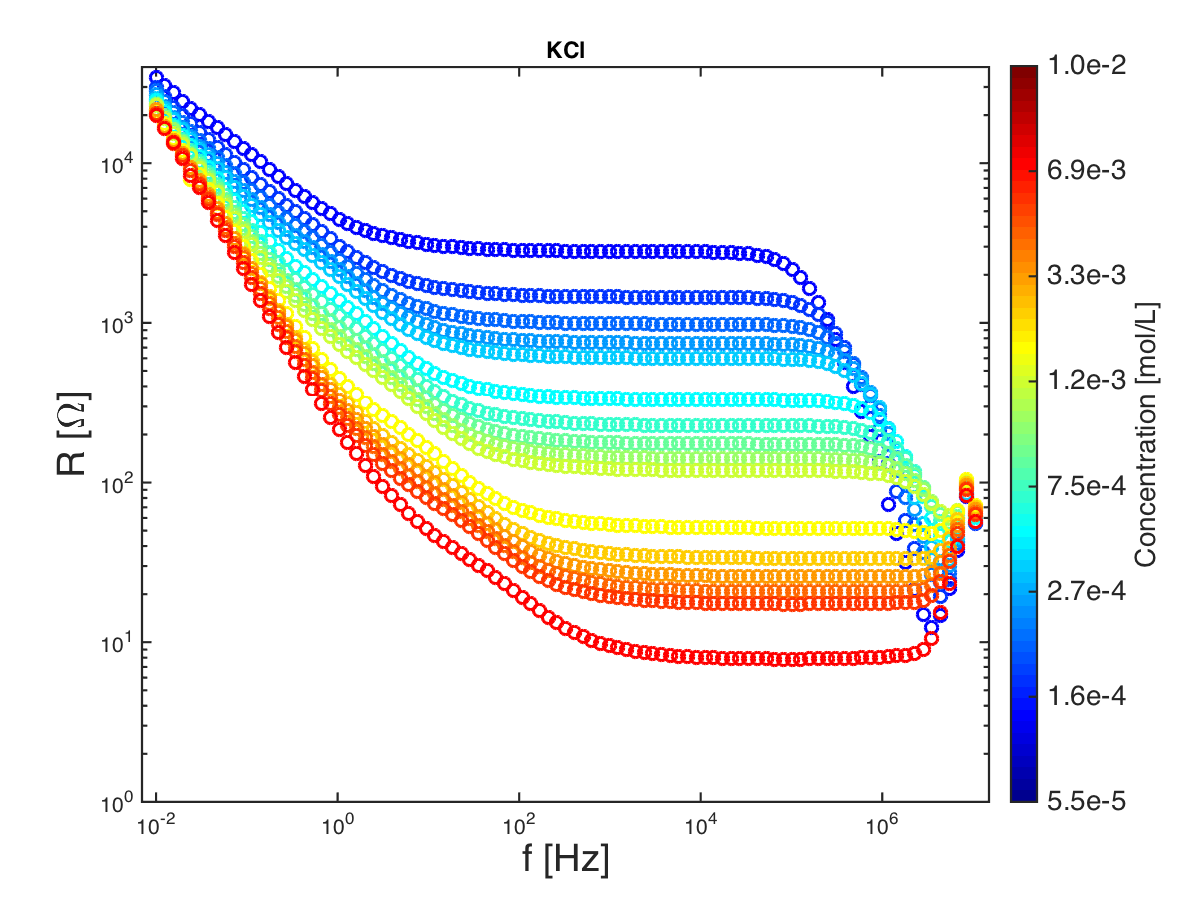}
		\caption{Resistance ($R$) vs. frequency ($f$) for the KCl solutions in a double logarithmic scale. Symbols and concentrations as in Fig.~\ref{NaClR}.}
		\label{KClR}
\end{figure}

\begin{figure}[htb]
	\includegraphics[width=0.75\textwidth]{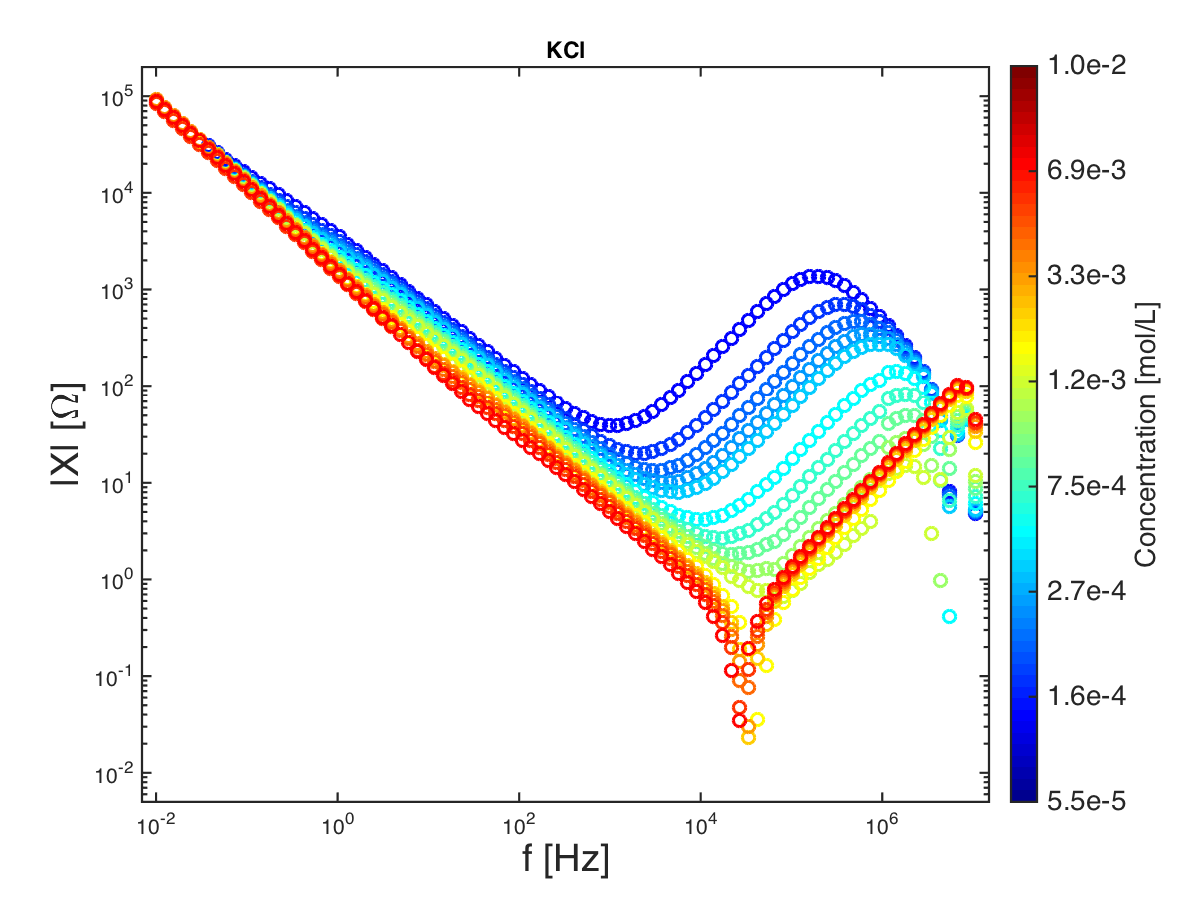}
		\caption{Absolute value of the reactance ($\left|X\right|$) vs. frequency ($f$) for the KCl solutions in a double logarithmic scale. Symbols and concentrations as in Fig.~\ref{NaClR}.}
		\label{KClX}
\end{figure}

In Figures~\ref{NaClR} -- \ref{KClX} we present in double logarithmic scaling the real, $R$, and imaginary, $X$, parts of the experimental data obtained by dielectric spectroscopy measurements, for the $16$ solutions of each salt. In particular, Fig.~\ref{NaClR} shows the resistence, $R$, versus frequency, $f$, spectra for each NaCl solution. Figure~\ref{NaClX} shows the corresponding spectra of the reactance absolute value, ($\left|X\right|$), versus frequency. $R$ and $X$ are measured in $\Omega$ and the frequency in $Hz$. Figure~\ref{KClR} shows $R$ versus $f$ for each KCl solution. Finally, Fig.~\ref{KClX} shows the corresponding spectra of $\left|X\right|$ versus $f$.

\end{document}